\documentclass[aps,prb,twocolumn,showpacs,superscriptaddress]{revtex4}
\usepackage{amsmath}
\usepackage{graphicx}
\usepackage{amssymb}
\usepackage{epsfig}
\usepackage{times}

\begin{document}

\title{U(1) spin liquids and valence bond solids in a large-$N$ three-dimensional Heisenberg model}
\author{Jean-S\'ebastien Bernier}
\author{Ying-Jer Kao}
\affiliation{Department of Physics, University of Toronto, Toronto,
Ontario, Canada M5S 1A7}
\author{Yong Baek Kim}
\affiliation{Department of Physics, University of Toronto, Toronto,
Ontario, Canada M5S 1A7}
\affiliation{School of Physics, Korea Institute for Advanced Study, Seoul 130-722, 
Korea}
\date{\today}

\begin{abstract}
We study possible quantum ground states of the Sp($N$) generalized Heisenberg model 
on a cubic lattice with nearest-neighbor {\it and} next-nearest-neighbor 
exchange interactions. The phase diagram is obtained in the large-$N$ limit and fluctuation
effects are considered via appropriate gauge theories. In particular, we find three U(1) spin 
liquid phases with different short-range magnetic correlations. These phases are characterized
by deconfined gapped spinons, gapped monopoles, and gapless ``photons''. 
As $N$ becomes smaller, a confinement transition from these phases to valence bond 
solids (VBS) may occur. This transition is studied by using duality and analyzing the resulting
theory of monopoles coupled to a non-compact dual gauge field; the condensation of the 
monopoles leads to VBS phases. We determine the resulting VBS phases emerging from
two of the three spin liquid states. On the other hand, the spin liquid state near $J_1 \approx J_2$ 
appears to be more stable against monopole condensation and could be a promising 
candidate for a spin liquid state in real systems.
\end{abstract}


\maketitle

\section{Introduction}
The search for quantum spin liquid states in frustrated magnets has long
been of great interest for condensed matter physicists. These exotic states
exhibit fractionalized quantum numbers and  may play 
important roles in understanding properties of  novel complex
materials such as high-$T_c$ cuprates. In the spin liquid states, 
there exist deconfined spinons with spin quantum number $S=1/2$, as opposed
to the more familiar magnon excitations in the N\'eel state, or the spin
triplet $S=1$ excitations in the paramagnetic spin-Peierls states.\cite{sachdevbook} Previous studies of
frustrated Heisenberg models on two dimensional lattices have revealed
possible quantum spin liquid phases corresponding to the deconfined phases of
$Z_2$ gauge theories.\cite{sachdev,chung01,bernier04,balents02}.
These phases are characterized by gapped spinons and gapped $Z_2$-vortices
that carry the $Z_2$ flux.\cite{rokhsar88,wen91,senthil,visontheory,moessner02,nayak01,demler02} 
Due to the gapped spectra of both spinons and $Z_2$-vortices, it is difficult
to observe an effect on measurable ``local'' quantities that distinguish the $Z_2$ spin liquid from 
a regular paramagnetic state. Theoretical proposals\cite{visontheory} have been
presented to observe the $Z_2$-vortices; however, to this date, experimental attempts have 
shown negative results.\cite{visonexp}

U(1) analogues of these phases, the so-called U(1) spin liquids, are not stable
due to instanton effects in two dimensions. 
On the other hand, deconfined phase of U(1) gauge theory can be stabilized
in three dimensions (3D)\cite{polyakov,gapless} and the resulting U(1) spin 
liquid states support gapped spinons, gapped monopoles, and gapless ``photons''.\cite{wen02,wen03}
Experimentally, these phases could be more easily observed since the gapless 
``photons'' would show up in the low-energy physical properties of the system
such as thermodynamics and energy/entropy transport. For example, an 
additional $T^3$ contribution to the specific heat from ``photons'' could be observed.        
There have been recent theoretical studies of three dimensional
models in which possible U(1) spin liquid phases are identified.\cite{huse04,hermele04,motrunich04} 
These models, however, are constructions in the limits where a large
degeneracy in the ground state manifold is first achieved by hand; for example, 
in Ref~\onlinecite{hermele04}, one needs to first take the easy-axis limit of the original 
Heisenberg model to facilitate further analysis. One desires to
obtain U(1) spin liquid phases without taking these special limits.

In this paper, we consider the Sp($N$) generalized three-dimensional frustrated 
Heisenberg model on a cubic lattice with nearest-neighbor(NN) $J_1$ and next-nearest-neighbor 
(NNN) $J_2$ interactions, using the bosonic Sp($N$) generalization 
of the physical spin SU($2$) $\cong$ Sp($1$) symmetry. 
Generalized spin operators can be
expressed  in terms of boson operators $b_i^{\dagger \alpha},b_{i\alpha}$ 
at each site $i$ where $\alpha =1, \dots, 2N $ labels the Sp($N$) index, 
and the constraint $n_b = b_i^{\dagger \alpha} b_{i\alpha}$ is imposed 
to fix the number of bosons per site ($n_b=2S$ for $N=1$ or the SU(2) limit).
This model has an advantage over other models as the spin-rotational
symmetry is retained, and there is a direct connection with the microscopic model.
We study the large-$N$ limit of this model and examine 
the finite-$N$ fluctuations via gauge theories. 
The phases obtained in the large-$N$ limit could also be relevant to physical 
systems whose microscopic Hamiltonians are ``near'' the 
parameter space of the original Hamiltonian. 
This idea, for example, has been used to explain the observed 
spin-correlations in the 2D spin-liquid state in Cs$_2$CuCl$_4$.\cite{chung03,coldea01,coldea03}
Moreover, recent advances of trapped atoms in optical lattices\cite{greiner02} allow one to construct 
atomic systems with enhanced symmetries like Sp($N$) or SU($N$);\cite{honerkamp03} large-$N$ theories
would provide useful information about the ground states of such systems.
  
At the mean-field level, we solve the Hamiltonian in the $N \rightarrow \infty$ limit
with $\kappa=n_b/N$ fixed. In general, smaller (larger) $\kappa$ corresponds 
to more (less) quantum fluctuations. As a result, magnetically long-range-ordered (LRO)
phases appear in the large $\kappa$ limit and quantum-disordered paramagnetic phases 
arise in the small $\kappa$ limit. Moreover, different LRO and paramagnetic phases 
show up depending on the ratio $J_2/J_1$.
We find three different U(1) spin liquid states obtained by quantum-disordering
LRO states with the ordering wave-vectors $(\pi,\pi,\pi),(0,\pi,\pi)$ and $(0,0,\pi)$.
These U(1) spin liquid phases are stable in the large-$N$ limit due to gapped
monopoles and still have short-range-ordered (SRO) spin correlations at the 
corresponding wave-vectors.

As $N$ becomes smaller, however, monopoles may condense
by closing the gap, and confined phases such as valence bond solids (VBS) 
may arise. We examine possible monopole condensation patterns and find
resulting VBS states that may emerge from the U(1) spin liquid phases. 
Different VBS phases are obtained from the U(1) spin liquid phases with
the SRO correlations at $(\pi,\pi,\pi)$ and $(0,0,\pi)$.
On the other hand, monopole condensation is not found in the U(1)
spin liquid with $(0,\pi,\pi)$ SRO in the simplest consideration of
monopole condensation. This may imply that one needs to go beyond
the approximation schemes we used in the analysis, or this U(1) spin liquid 
phase near $J_1 \approx J_2$ is more stable against monopole 
condensation. If the latter is the case, this may be a more promising
candidate for a U(1) spin liquid state in real materials.

The rest of the paper is organized as follows. In Sec.~\ref{formalism}, we introduce
the formalism used in this work. In Sec.~\ref{meanfield}, the large-$N$ mean-field phase diagram is
explained. In Sec.~\ref{spinlqd}, fluctuation effects about the mean-field states are 
considered via U(1) gauge theory and its dual theory of monopoles coupled
to a noncompact U(1) gauge field.
Here, three U(1) spin liquid states are also identified. 
In Sec.~\ref{vbs}, possible VBS phases that may arise at finite-$N$ are 
obtained using the dual theory of monopoles with frustrated hopping
on the dual lattice. We conclude in Sec.~\ref{conclusion}.

\section{Formalism\label{formalism}}

The Hamiltonian for our model can be written as 

\begin{equation}
H = J_1 \sum_{\left < ij \right >} \mathbf{S}_i \cdot \mathbf{S}_j
+ J_2 \sum_{\left < \left < ij \right > \right >} 
\mathbf{S}_i \cdot \mathbf{S}_j ,
\label{hamil}
\end{equation}
where $\mathbf{S}_i$ are $S=1/2$ operators at site $i$.
Here $J_1 > 0$ is the antiferromagnetic exchange coupling
on the NN links, and 
$J_2 > 0$ on the
NNN links.
The proper description of a frustrated antiferromagnet requires that
all spins transform under the same representation of a group, and 
two spins can combine to form a singlet. This requirement leads one to
consider the generalization of the physical spin SU($2$) $\cong$ Sp($1$) 
symmetry to Sp($N$) and study the quantum ground states of the 
Sp($N$) Hamiltonian in the large-$N$ limit.\cite{sachdev}

As the first step toward Sp($N$) generalization, one rewrites the
SU($2$) spin operators using their bosonic representation,
$\mathbf{S}_i = \frac{1}{2}b_i^{\dagger \alpha}
\mathbf{\sigma}_\alpha^{\beta}b_{i\beta}$,
where $\alpha, \beta = \uparrow, \downarrow$ labels two possible spin
states of each boson, $b_{i \alpha}$, and the constraint
$n_b = b_i^{\dagger \alpha} b_{i\alpha} = 2S$
must be imposed at each site. The Heisenberg Hamiltonian, an additive 
constant aside, is then given by

\begin{equation}
H = -\frac{1}{2}\sum_{ij}J_{ij}(\epsilon_{\alpha \beta}
b_i^{\dagger \alpha}b_j^{\dagger \beta})
(\epsilon^{\gamma \delta} b_{i\gamma} b_{j\delta})
\label{newhamil}
\end{equation}
where $J_{ij} = J_1$ is on the NN links, and
$J_{ij} = J_2$ is on the NNN links. Here $\epsilon_{\alpha \beta}$
is the antisymmetric tensor of SU(2).
One can then generalize this expression to Sp($N$) by formally
introducing $N$ flavors of bosons on each site, and by changing the
constraint to $n_b = b_i^{\dagger \alpha} b_{i\alpha} = 2NS$, where
$\alpha = 1, ..., 2N$ is the Sp($N$) index. For the physical case, $N=1$,
$S$ takes half-integer values. Following these transformations,
the Sp($N$) Hamiltonian becomes
\begin{equation}
H = -\frac{1}{2N}\sum_{ij}J_{ij}(\mathcal{J}_{\alpha \beta}b_i^{\dagger \alpha}
b_j^{\dagger \beta})(\mathcal{J}^{\gamma \delta} b_{i\gamma} b_{j\delta}),
\label{spnhamil}
\end{equation}
where $\mathcal{J}^{\alpha \beta} = \mathcal{J}_{\alpha \beta} =
- \mathcal{J}_{\beta \alpha}$ is the generalization of the antisymmetric tensor
of SU(2); it is a $2N \times 2N$ matrix that contains $N$ copies of $\epsilon$
along its center block diagonal and vanishes elsewhere.\cite{sachdev}

In the $N \rightarrow \infty$ limit at a fixed boson density per flavor,
$n_b/N = 2S = \kappa$, we obtain to leading order in $1/N$ a mean field theory 
for $S = \kappa/2$. The fluctuations about the mean field solution give rise 
to a gauge theory.\cite{sachdev} The mean-field phase diagram at $N \rightarrow 
\infty$ is shown in Fig.~\ref{fig:pd} as a function of $J_2/(J_1+J_2)$ and $1/\kappa$.
At large values of $S$, various magnetically long-range-ordered (LRO)
phases appear and are represented by the ordering wave-vector 
${\bf q}=(q_1,q_2,q_3)$.
The short-range-ordered (SRO) phases at small values of $S$
correspond to quantum-disordered phases with
short-range equal-time spin correlations enhanced at the
corresponding wave-vectors.

\section{Mean Field Phase Diagram \label{meanfield}}

The mean-field theory can be obtained by decoupling the quartic 
boson interactions in $\mathcal{S}$ using Hubbard-Stratonovitch 
fields $Q_{ij}=-Q_{ji}$ directed along the lattice links.
The effective action then contains the terms
\begin{equation}
\mathcal{S}=\int{d\tau \sum_{i>j}\frac{J_{ij}}{2}\left[N \lvert Q_{ij}\lvert^2
-Q_{ij}^{*}\mathcal{J}_{\alpha\beta}b_i^{\alpha}b_j^{\beta}+c.c.\right]+\cdots},
\label{action}
\end{equation}
where $\tau$ is the imaginary time and the ellipses represent standard terms which
impose the canonical boson commutation relations and the constraint.\cite{sachdev}
At the saddle point of the action, we get
\begin{equation}
\left < Q_{ij} \right > = \frac{1}{N} \left < \mathcal{J}^{\alpha\beta}
b_{i\alpha}^{\dagger}b_{j\beta}^{\dagger} \right >.
\end{equation}
The one-site unit cell of the cubic lattice
has nine of these $Q_{ij}$ fields. For larger values of $S=\kappa/2$, the dynamics of 
$\mathcal{S}$ leads to the condensation of the $b_i^{\alpha}$ bosons and one 
obtains a nonzero
value of
\begin{equation}
\left < b^{\alpha}_i \right > = x^{\alpha}_i.
\end{equation}
This corresponds to a magnetically-ordered phase.

\begin{figure}
\includegraphics[height=6cm,width=8.5cm]{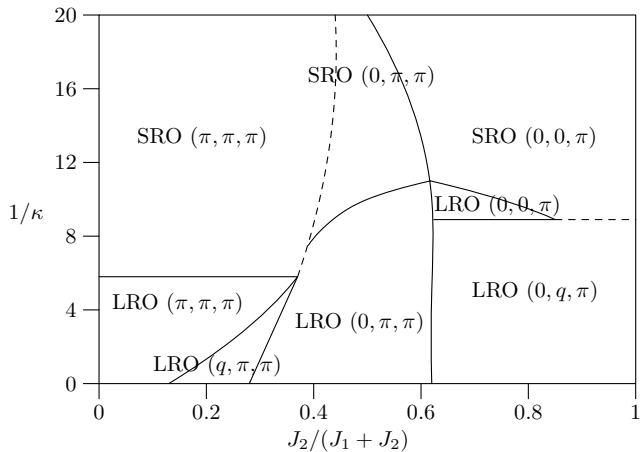}
\caption{\label{fig:pd} Large-$N$ phase diagram of the Sp($N$) cubic
lattice model with first and second-nearest neighbor interactions 
as a function of $J_2/(J_1+J_2)$ and $1/\kappa$. The LRO phases break
spin-rotation symmetry. The spin order is collinear and commensurate in 
the $(\pi,\pi,\pi)$, $(0,\pi,\pi)$ and $(0, 0, \pi)$ LRO phases while it 
is helical and incommensurate in the $(q, \pi, \pi)$ and $(0, q, \pi)$  
LRO phases. The SRO phases preserve spin-rotation invariance. Notice
that the $(q, \pi, \pi)$ and $(0, q, \pi)$ LRO phases do not have SRO 
counterparts. [Dashed line: first order transition. Solid line:
continuous transition.]}
\end{figure}

The large-$N$ limit of $\mathcal{S}$ is taken for a fixed value 
of $\kappa = n_b/N = 2S$ and, depending on the ratio $J_1/J_2$ and the value 
of $\kappa$, the ground state of $\mathcal{S}$ at $T=0$ can either break 
the global Sp($N$) symmetry and posses magnetic LRO
or be Sp($N$) invariant with only SRO. We optimized the ground state energy
with respect to variations in $\left<Q_{ij}\right>$ and $x_i^{\alpha}$ 
for different values of $J_2/J_1$ and $\kappa$. 
We also found that each saddle point may be described
by a purely real $\left<Q_{ij}\right>$.
The phase diagram is summarized in Fig.~\ref{fig:pd}. 
Notice the transition between different phases are in general second-order
(solid lines), with the exception of two first-order transitions (dashed
lines). The various magnetically  ordered and paramagnetic 
 phases are described in details as follows.

\subsection{Magnetically ordered phases}

The magnetically ordered phases are characterized by the finite condensate
$x^{\alpha}_i \not= 0$. In this model, we find the following LRO states.

\subsubsection{$(\pi,\pi,\pi)$ LRO state}

\begin{figure}
\includegraphics[width=3.8cm]{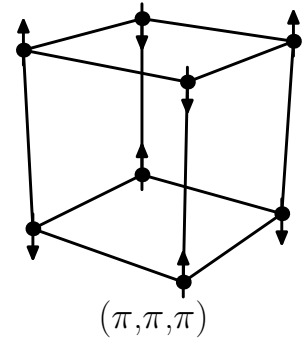}
\caption{\label{fig:pipipi} $(\pi, \pi, \pi)$ phase. All NN 
         bonds have nonzero values.} 
\end{figure}

This is a long-range-ordered state in which $\langle\mathbf{S}_i\rangle$ 
is collinearly polarized in opposite directions on two
inter-penetrating cubic sublattices (Fig.~\ref{fig:pipipi}). 
A gauge can be chosen in which the expectation values of 
link variables are nonzero and equal on horizontal 
and vertical links, while the values on diagonal links
are zero.

\subsubsection{Helical $(q, \pi, \pi)$ LRO state}

\begin{figure}
\includegraphics[width=3.8cm]{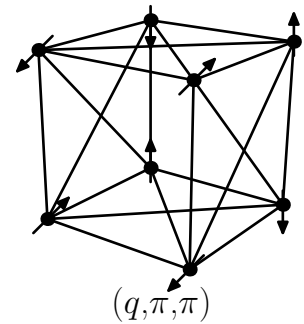}
\caption{\label{fig:qpipi} $(q, \pi, \pi)$ phase. All NN 
         bonds have nonzero values and NNN 
         bonds are nonzero on four of the six possible faces.} 
\end{figure}

This helically-ordered phase is characterized by nonzero values of
$\langle Q_{ij}\rangle$ on all NN links and on four
faces of the cube (each face has two NNN 
links (Fig.~\ref{fig:qpipi})). In the appropriate gauge, two of the
three possible NN bond directions present the same
expectation value. Due to the three rotation axes of the cube,
there are three possible choices for the nonequivalent NN
link direction: $(q,\pi,\pi), (\pi,q,\pi), (\pi,\pi,q)$. 
This choice dictates on which four faces
will lie the NNN bonds with nonzero
expectation value. It is worth mentioning that all
nonzero diagonal bonds have the same strength. Also, this phase 
has a long-range incommensurate spin order and the spin structure 
factor peaks at the incommensurate wave vector $(q, \pi, \pi)$.

\subsubsection{$(0, \pi, \pi)$ LRO state}

\begin{figure}
\includegraphics[width=3.8cm]{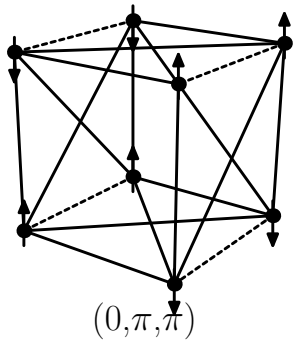}
\caption{\label{fig:0pipi} $(0, \pi, \pi)$ phase. NN 
         bonds are nonzero in two directions and NNN 
         bonds are nonzero on four out of the six possible faces.
         [Note: dashed lines are only a guide to the eye.]} 
\end{figure}

This magnetically ordered phase has nonzero bond expectation values
for two of the three possible NN bond 
directions. Moreover, NNN links are nonzero and 
equal on four faces of the cube (Fig.~\ref{fig:0pipi}). Once again,
there are three possible choices of direction for the zero NN
bond since these three configurations are interchangeable under 
rotation around the Cartesian axes of the cube.

\subsubsection{$(0, 0, \pi)$ LRO state}

\begin{figure}
\includegraphics[width=3.8cm]{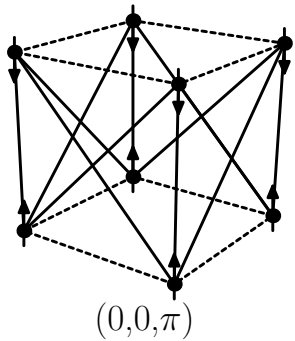}
\caption{\label{fig:00pi} $(0, 0, \pi)$ phase. Only
         NN bonds in one direction can be nonzero
         while NNN
         links are nonzero on four of the six faces.
         [Note: dashed lines are only a guide to the eye.]}
\end{figure}

This phase is characterized by nonzero NN bonds in
only one of the three possible directions while NNN
links are nonzero on four of the six faces (Fig.~\ref{fig:00pi}).
There are three possible choices for the nonzero 
NN bond direction.

\subsubsection{Helical $(0, q, \pi)$ LRO state}

\begin{figure}
\includegraphics[width=3.8cm]{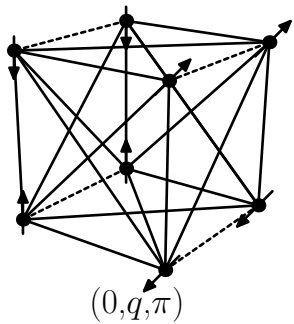}
\caption{\label{fig:0qpi} $(0, q, \pi)$ phase. NN 
         bonds are nonzero in two directions and decrease 
         continuously as $J_1$ decreases. All NNN 
         bonds are nonzero. [Note: dashed lines are only a 
         guide to the eye.]}
\end{figure}

This helically-ordered phase presents nonzero expectation
values for all NNN bonds, and two of
the NN bond directions are also nonzero  
(Fig.~\ref{fig:0qpi}). Diagonal bonds do not exhibit the same strength, and
the NN bond values continuously decrease and 
reach zero at $J_1=0$. Moreover, this phase exhibits a 
long-range incommensurate spin order and the spin structure
factor peaks at the incommensurate wave vector $(0, q, \pi)$.

\subsection{Paramagnetic phases}

The paramagnetic phases are characterized by the vanishing condensate 
$x^{\alpha}_i = 0$ and a gapped spinon excitation spectrum. We find three
distinct SRO states.

\subsubsection{$(\pi,\pi, \pi)$ SRO state}

This state is obtained by quantum disordering the 
$(\pi, \pi, \pi)$ LRO state. The expectation values of $Q_{ij}$
have similar structures as those of its ordered counterpart.
In this state, all spin excitations are gapped.

\subsubsection{$(0, \pi, \pi)$ SRO state}

Once again, this state is obtained by quantum disordering the 
corresponding LRO state. The expectation values of $Q_{ij}$
have similar structures as those of its ordered counterpart.
Although all spin excitations are gapped in this state,it is not 
symmetric under $\pi/2$ rotation around two of the Cartesian
axes (y and z).

\subsubsection{$(0, 0, \pi)$ SRO state}

This state is the quantum disordered counterpart of the
$(0, 0, \pi)$ LRO phase. The expectation values of the four NN links, 
not strictly null, are very small and will reach
zero as $J_1$ decreases. Even though all spin excitations are
gapped in this state, it is not symmetric under $\pi/2$ rotation 
around two of the Cartesian axes (x and y).


\subsection{Analysis beyond mean-field for the paramagnetic phases}

It is important to note that these three 
SRO phases possess bipartite lattice structure.
To get a clearer picture, we label ``up-spin'' sites,
where bosons carry a $+1$ ``charge'', with {\it A}, and
``down-spin'' sites, where bosons carry a $-1$ ``charge'',
with {\it B}. Since all non-zero
$Q$-fields are on bonds between {\it A} and {\it B} sublattices, 
these fields will form ``charge''-zero objects. In other words, the 
mean field parameter $\langle Q_{ij}\rangle$ is invariant 
under a global U(1) gauge transformation:
\begin{align}
&b_{i}\rightarrow b_{i} e^{i\phi},  \nonumber \\
&b_{j}\rightarrow b_{j} e^{-i\phi}, \nonumber \\
&Q_{ij}\rightarrow Q_{ij},
\end{align}
where $i$ $\in$ {\it A} and $j$ $\in$ {\it B}.
Thus, unlike the $Z_2$ spin liquid cases in two-dimensions,
the U(1) symmetry of the action is intact.
These phases would exhibit low-energy gapless
U(1) gauge excitations (``photons'').

However, to correctly capture the structure of these phases beyond 
the mean-field level, we also need to consider the contribution from topologically
non-trivial gauge-field configurations, namely monopoles.
In particular, the interference effects between monopole events due to 
Berry phases lead to different quantum ground states.
In general, two possible classes of paramagnetic ground states exist. 
In the first class of paramagnets, none of the symmetries of the Hamiltonian 
is broken. The spins are paired into valence bond singlets which strongly resonate 
between a large number of possible valence bond configurations. 
Such a state is a spin liquid or a resonating valence bond liquid. 
In the second class of paramagnets, the valence bond singlets
spontaneously crystallize into a configuration which necessarily 
breaks a lattice symmetry. This configuration is a valence bond solid. 
We can distinguish these two classes of paramagnets using the 
expectation value of the monopole field as our order parameter. 
Monopoles are gapped in a spin liquid, and condense in a valence
bond solid. 

Consequently, in the next sections, we investigate under
which conditions the three paramagnetic phases are 
spin liquids or valence bond solids. First, in section 
\ref{spinlqd}, we write down the action for such a system
and find in which limit we get spin liquids. Then, 
in section \ref{vbs}, we study the effect of monopole 
condensation and its consequence on the structure 
of the SRO phases.

\section{Monopole action, Berry phases, and U(1) spin liquids \label{spinlqd}}

The SRO phases described above have gapped spinons, hence we can safely
integrate out spinons and study the resulting compact U(1) gauge 
theory.\cite{sachdev90}
This analysis was used successfully in (2+1)D to study antiferromagnets 
on the square\cite{sachdev92}, Shastry-Sutherland\cite{chung01} ,
and checkerboard\cite{bernier04} lattices.
Thus, our starting point is the action of a compact $U(1)$ 
gauge theory in (3+1)D on a space-time lattice in the 
presence of a Berry phase term. Physically, the Berry phase term 
can be assimilated to a static 
background gauge charge of strength $\pm1$ on the sites 
of the spatial cubic lattice. This action is given by
\begin{align}
\mathcal{Z}_A &= \int_{0}^{2\pi} \mathcal{D}[A_{j\mu}] \exp 
\Big(K \sum_{j,\mu < \nu} 
\cos(\Delta_{\mu}A_{j\nu} - \Delta_{\nu}A_{j\mu}) \nonumber \\
&~~~~~~ - i 2S \sum_{j} \eta_{j} A_{j\tau} \Big)
\label{cos_act}
\end{align}
where $j$ denotes the sites
of the space-time lattice, and the cosine term represents the
conventional Maxwell action for a compact U(1) gauge theory. It
is the simplest local term consistent with the gauge symmetry
$A_{j\mu}\rightarrow A_{j\mu}-\phi_j-\phi_{j+\mu}$ and it is
periodic under $A_{j\mu}\rightarrow A_{j\mu}+2\pi$. The
last term is the crucial Berry phase leading to 
large cancellations between different paths. 
The static background gauge charges,
$\eta_j$, can be $\pm 1$ depending on the underlying
magnetic structure.

In (2+1)D, the presence of an alternating static background 
charge lead, in confined paramagnetic phases, to broken
translational symmetry. Here, our aim is to extend this 
analysis to (3+1)D paramagnets by performing duality
transformations on $\mathcal{S}$. Neglecting the Berry phase
term, the remaining compact U(1) gauge theory
becomes a theory of point-like monopoles coupled to the dual
non-compact U(1) gauge field. However, the presence of the
background charge complicates the situation. 
The Berry phase term corresponds to a dual magnetic flux emanating 
from the center of each dual cube. This flux alternates in sign 
following a pattern particular to each SRO phase. Thus, 
we expect to get, after a series of dual transformations on 
$\mathcal{S}$, a theory of monopoles with frustrated hopping 
coupled to a non-compact gauge field. In the remainder of this
section, we first explicitly show how to obtain the
aforementioned monopole action. Then we present the conditions
under which these SRO phases are U(1) spin liquids.

\subsection{Duality transformations: path towards ``frustrated'' 
monopole action}

As a first step, we replace the cosine 
interaction by a Villain sum over periodic Gaussians such that
\begin{align}
&\exp\Big(K \cos(\Delta_{\mu}A_{j\nu} - \Delta_{\nu}A_{j\mu})\Big) \nonumber \\ 
& \rightarrow \sum_{q_{\overline{j}\mu\nu}}
\exp\Big(-\frac{K}{2} (\Delta_{\mu}A_{j\nu} - \Delta_{\nu}A_{j\mu}
- 2\pi q_{\overline{j}\mu\nu})^2\Big),
\end{align}
where $\overline{j}$ denotes the sites of the dual lattice.
We then rewrite this expression using Poisson resummation
formula
\begin{align}
&\exp\Big(K \cos(\Delta_{\mu}A_{j\nu} - \Delta_{\nu}A_{j\mu})\Big) \nonumber \\ 
& \rightarrow \sum_{f_{j\mu\nu}} \int_{-\infty}^{\infty} 
\mathcal{D}[\phi_{\overline{j}\mu\nu}] \nonumber \\
&\exp\Big(-\frac{K}{2}(\Delta_{\mu}A_{j\nu} - \Delta_{\nu}A_{j\mu} 
-2\pi\phi_{\overline{j}\mu\nu})^2 
+ i 2\pi f_{j\mu\nu} \phi_{\overline{j}\mu\nu}\Big),
\end{align}
where $f_{j\mu\nu}$ lives on the links of the direct lattice.
Following the change of variable $\xi_{\overline{j}\mu\nu}= 
\Delta_{\mu}A_{j\nu} - \Delta_{\nu}A_{j\mu} -2\pi\phi_{\overline{j}\mu\nu}$, 
we perform Gaussian integrals over the intermediate fields 
$\xi_{\overline{j}\mu\nu}$ and obtain
\begin{align}
&\exp\Big(K \cos(\Delta_{\mu}A_{j\nu} - \Delta_{\nu}A_{j\mu})\Big) \nonumber \\ 
&\rightarrow \sum_{f_{j\mu\nu}} 
\exp\Big(-\frac{f_{j\mu\nu}^2}{2K} + i f_{j\mu\nu}(\Delta_{\mu}A_{j\nu} 
- \Delta_{\nu}A_{j\mu})\Big).
\end{align}
Upon performing these transformation on (\ref{cos_act}),
the resulting partition function is
\begin{align}
\mathcal{Z}_A &= \int_{0}^{2\pi} \mathcal{D}[A_{j\mu}] 
\sum_{f_{j\mu\nu}} \exp\Big(\sum_{j,\mu<\nu} \frac{- f_{j\mu\nu}^2 }{2K} \nonumber \\
&~~~~~ +i f_{j\mu\nu}(\Delta_{\mu}A_{j\nu} - \Delta_{\nu}A_{j\mu})
-i2S\sum_j\eta_j A_{j\tau}\Big),
\label{transfaction}
\end{align}
where the integer-valued field $f_{j,\mu\nu}$ can be interpreted 
as an analogue of the electromagnetic field tensor. For completeness, it should be noted 
that we dropped in Eq.~(\ref{transfaction}) an insignificant overall normalization 
constant. 

We can then perform on this more amenable expression the integrals
over $A_{j\mu}$, and we find (for $S=1/2$) the condition
\begin{equation}
\Delta_{\nu}f_{j\mu\nu}=\eta_j\delta_{\mu\tau}.
\end{equation}
This condition can take the form 
\begin{align}
& \Delta_a e_{ja} = \eta_j \nonumber \\
& \epsilon_{abc} \Delta_b b_{\overline{j}c} - \Delta_{\tau}e_{ja} = 0
\end{align}
where $e_{ja}$ is the ``electric field'' on the links of the
direct lattice and $b_{\overline{j}a}$ is the ``magnetic field'' on the 
links of the dual lattice. Upon solving these two equations, we obtain
\begin{align}
&e_{ja} =  e^{0}_{ja} + \epsilon_{abc}\Delta_b h_{\overline{j}c} \nonumber \\
&b_{\overline{j}a} = \Delta_{\tau}h_{\overline{j}a}
+\Delta_a\psi_{\overline{j}}.
\end{align}
Thus, $\Delta_a e_{ja}^{0}=\eta_j$, and we can rewrite the 
partition function in terms of $e_{ja}$ and  $b_{\overline{j}a}$
\begin{align}
\mathcal{Z}_{A} &= \sum_{\{e_{ja},b_{\overline{j}a}\}}
\exp\Big(-\frac{1}{2K} \sum_{j,\overline{j},a} (e_{ja}^2 
+b_{\overline{j}a}^2)\Big)\nonumber \\
&= \sum_{\{h_{\overline{j}a}, \psi_{\overline{j}a}\}} \exp\Big(-\frac{1}{2K}\sum_{j,\overline{j},a} 
\big((e_{ja}^{0}+\epsilon_{abc}\Delta_b h_{\overline{j}c})^2 \nonumber \\ 
&~~~~~~ + (\Delta_{\tau} h_{\overline{j}a}+
\Delta_a\psi_{\overline{j}})^2\big)\Big).
\label{act_eb}
\end{align}
The physical properties of the partition function become
clearer by parameterizing the ``frustration'' $e^{0}_{ja}$ into a
curl in terms of a new fixed field $Y_{\overline{j}c}$. Thus,
\begin{equation}
e^{0}_{ja}= \epsilon_{abc}\Delta_b Y_{\overline{j}c}.
\label{param}
\end{equation}
Inserting Eq.~(\ref{param}) into Eq.~(\ref{act_eb}), we can now write
the partition function in the following form:
\begin{align}
\mathcal{Z}_{A} &= \sum_{\{ h_{\overline{j}a}, \psi_{\overline{j}a}\}} \exp\Big(-\frac{1}{2K}\sum_{j,\overline{j},a}
((\epsilon_{abc}\Delta_b(h_{\overline{j}c}+Y_{\overline{j}c}))^2 \nonumber \\
&~~~~~~ + (\Delta_{\tau}h_{\overline{j}a}+\Delta_a\psi_{\overline{j}})^2)\Big).
\label{act_param}
\end{align}
To write Eq.~(\ref{act_param}) in a simpler form, we define a new dual field
$L_{\overline{j}a}=h_{\overline{j}a}+Y_{\overline{j}a}$, and make the gauge
choice $\psi_{\overline{j}}=-L_{\overline{j}\tau}$. As a result, our theory
is now given by
\begin{equation}
\mathcal{Z}_{A} = \sum_{\{L_{\overline{j}\rho}\}} \exp \Big(-\frac{1}{2K}
\sum_{\overline{j},\rho<\sigma}(\Delta_{\sigma}L_{\overline{j}\rho}
-\Delta_{\rho}L_{\overline{j}\sigma})^2 \Big).
\end{equation}

We then promote the integer-valued field $L_{\overline{j}\rho}$ 
to real values with appropriate conditions. 
To do so, we first use the Poisson resummation formula 
to soften the integer constraint on $L_{\overline{j}\rho}$. The cost 
of this procedure is the introduction of a new integer-valued
field $J_{\overline{j}\rho}^{(m)}$, the monopole
four-current. After shifting the real field by 
$L_{\overline{j}\rho} \rightarrow L_{\overline{j}\rho}+
Y_{\overline{j}\rho}$, we obtain
\begin{align}
&\mathcal{Z}_{A} =\nonumber \\ 
&\sum_{\{J_{\overline{j}\rho}^{(m)}\}} 
\int_{-\infty}^{\infty} \mathcal{D}[L_{\overline{j}\rho}]
\exp \Big(-\frac{1}{2K}
\sum_{\overline{j},\rho<\sigma}(\Delta_{\sigma}L_{\overline{j}\rho}
-\Delta_{\rho}L_{\overline{j}\sigma})^2 \nonumber \\
&~~~~~~ -2\pi i \sum_{\overline{j},\rho} J_{\overline{j}\rho}^{(m)}
(L_{\overline{j}\rho}+Y_{\overline{j}\rho})\Big).
\end{align}
Then, to control the fluctuations of $J_{\overline{j}\rho}^{(m)}$,
we add the mass term
\begin{equation}
S_{fugacity}=-\frac{\ln(\lambda_m)}{2}\sum_{\overline{j},\rho}(J_{\overline{j}\rho})^2.
\end{equation}
Moreover, to explicitly make the the dual theory gauge invariant and to
enforce monopole current conservation, we add a second term given by
\begin{equation}
\int_{0}^{2\pi}\mathcal{D}[\theta_{\overline{j}}^{(m)}]
\exp\left(2i\pi\theta_{\overline{j}}^{(m)}\Delta_{\rho}
J_{\overline{j}\rho}^{(m)}\right),
\end{equation}
where $\theta_{\overline{j}}^{(m)}$ is a $U(1)$ field existing
on each site of the dual lattice. 
Note that $\exp(i\widetilde{\theta}^{(m)}_{\overline{j}})$ 
corresponds to the monopole creation operator and $\lambda_m$ is 
the monopole fugacity.

Following these two transformations, 
the partition function is now given by
\begin{align}
\mathcal{Z}_{A} &= \sum_{\{J_{\overline{j}\rho}^{(m)}\}} 
\int_{-\infty}^{\infty} \mathcal{D}[L_{\overline{j}\rho}]
\int_{0}^{2\pi} \mathcal{D}[\theta_{\overline{j}}^{(m)}]
~~e^{-\mathcal{S}}, 
\end{align}
\begin{align}
\mathcal{S} &= \frac{1}{2K}\sum_{\overline{j},\rho<\sigma}
(\Delta_{\sigma}L_{\overline{j}\rho}-\Delta_{\rho}L_{\overline{j}\sigma})^2 \nonumber \\
&~~~-\sum_{\overline{j}\rho}\Big(\frac{\ln(\lambda_{m})}{2}
(J_{\overline{j}\rho}^{(m)})^2 \nonumber \\
&~~~~~~+2i\pi J_{\overline{j}\rho}^{(m)}(\Delta_{\rho}\theta_{\overline{j}}^{(m)} 
- L_{\overline{j}\rho} - Y_{\overline{j}\rho})\Big).
\end{align}
Finally, upon summing over $J_{\overline{j}\rho}^{(m)}$, and rescaling all fields
by $2\pi$, the final action reads
\begin{align}
\mathcal{Z}_{A} &= 
\int_{-\infty}^{\infty} \mathcal{D}[\widetilde{L}_{\overline{j}\rho}]
\int_{0}^{2\pi} \mathcal{D}[\widetilde{\theta}_{\overline{j}}^{(m)}]
~~e^{-\mathcal{S}}, 
\end{align}
\begin{align}
\mathcal{S} &= \frac{1}{8K\pi^2}\sum_{\overline{j},\rho<\sigma}
(\Delta_{\sigma}\widetilde{L}_{\overline{j}\rho}-\Delta_{\rho}
\widetilde{L}_{\overline{j}\sigma})^2 
\nonumber \\
&~~~-\lambda_m \sum_{\overline{j}\rho} 
\cos\big(\Delta_{\rho}\widetilde{\theta}_{\overline{j}}^{(m)} 
- \widetilde{L}_{\overline{j}\rho} - \widetilde{Y}_{\overline{j}\rho}\big).
\end{align}
It is important to note that, since the static gauge charge at the center 
of each dual cube is $\pm1$, the quantity 
$\epsilon_{abc}\Delta_b \widetilde{Y}_{\overline{j}c}$,
interpreted as fluxes through the faces of the dual cubes, is 
defined modulo $2\pi$.

To summarize our result, the duality transformations on $\mathcal{S}$ yield
a theory of monopoles with frustrated hopping on the dual lattice, coupled to 
a dual non-compact gauge field $\widetilde{L}_{\overline{j}\rho}$. 
The ``frustration'' is encoded in the field $\widetilde{Y}_{\overline{j}\rho}$ 
and arises from the Berry phase term. 

\subsection{U(1) spin liquids}

From the final expression of the dual action, we see that, for $\lambda_m$ 
close to zero, the monopole field is gapped, and the non-compact
dual gauge field is free. Thus, the resulting phase has gapped spinons,
has gapped monopoles, has gapless photons, and respects all the lattice symmetries.
This phase is a $U(1)$ spin liquid. Thus, for small fugacity, all three SRO 
phases can be spin liquids. It is interesting to note that the sole inclusion
of the NNN interaction in the cubic lattice has allowed
us to find three neighboring spin liquid phases in the parameter space.

\section{Valence bond solids \label{vbs}}

When $\lambda_m$ becomes of order one, monopole 
condensation gives rise to phases with broken symmetries. In this
section, we study different valence bond configurations that may arise from 
the three SRO phases. In order to achieve this, we first
ignore the fluctuations of the dual gauge field 
$\widetilde{L}_{\overline{j}\rho}$. Once the structure of the resulting 
slow fluctuation description is known, we may restore the 
field $\widetilde{L}_{\overline{j}\rho}$. Hence, 
we begin our study with the following action
\begin{equation}
\mathcal{S} = -\lambda_m \sum_{\overline{j}\rho} 
\cos\big(\Delta_{\rho}\widetilde{\theta}_{\overline{j}}^{(m)} 
- \widetilde{Y}_{\overline{j}\rho}\big).
\end{equation}
We then rewrite the cosine term using exponentials, and promote $\tau$ to
a continuous variable. The action now takes the form:
\begin{align}
\mathcal{S} =& -\frac{\lambda_{m}}{2}\int d\tau
\Big[\sum_{R}(e^{i\Delta_{\tau}\widetilde{\theta}_{R}^{(m)}}
+c.c.) \nonumber \\
~~~~&+\sum_{\left<RR'\right>}(
e^{-i\widetilde{Y}_{RR'}}e^{i(\widetilde{\theta}_{R'}^{(m)}-
\widetilde{\theta}_{R}^{(m)})}
+c.c.)\Big],
\end{align}
where ${\bf R}$ labels the sites on the cubic lattice 
dual to the original lattice of spins. The sites of 
the dual lattice are at the centers of the
cubes defined by the planes of the original lattice, 
and ${\bf R}=(x,y,z)$ with $x$, $y$ and $z$ integers 
in units of the lattice constant ($a=1$). 
Then, assuming 
that $\widetilde{\theta}_{R}^{(m)}$ is a slowly-varying 
function of time, we find
\begin{equation}
e^{i\Delta_{\tau}\widetilde{\theta}_{R}^{(m)}}
+c.c. \approx 2 - |\partial_\tau \widetilde{\theta}_{R}^{(m)}|^2
=2-|\partial_\tau\Phi_R|^2,
\label{approx}
\end{equation}
where $\Phi_R^{\dagger}= e^{i\widetilde{\theta}_{R}^{(m)}}$ is the monopole creation
operator.
Consequently, by dropping the constant term from Eq.~(\ref{approx}) 
and adding the potential 
$V(|\Phi_R|^2)=r_0|\Phi_R|^2+\mu_0|\Phi_R|^4+...$, 
we obtain the continuous-time soft-spin action for the 
frustrated XY model
\begin{align}
\mathcal{S}=&\int d\tau \Big(\sum_{R} |\partial_\tau\Phi_R|^2-
\sum_{\left< RR'\right>} (t_{RR'}\Phi^{*}_{R}\Phi_{R'}+
c.c.) \notag \\
& +\sum_{R}V(|\Phi_R|^2)\Big),
\label{sftspnaction}
\end{align}
where the frustration is encoded in the 
monopole hopping amplitudes $t_{RR'}=te^{i\widetilde{Y}_{RR'}}$ corresponding 
to the fluxes through the dual plaquettes. These flux values depend 
on the fixed field $\eta_j$ which distinguishes the different SRO phases, 
and encapsulates the spin staggering in the local collinear 
order. It has the values 
\begin{equation}
\eta_j = \left\{ \begin{array}{lc} 
(-1)^{j_z} &~\mbox{~~for $(0,0,\pi)$ SRO} \\
(-1)^{j_y + j_z} &~\mbox{~~for $(0,\pi,\pi)$ SRO} \\ 
(-1)^{j_x + j_y + j_z} &~\mbox{~~for $(\pi,\pi,\pi)$ SRO}.
\end{array} \right.
\label{etaj}
\end{equation}
In the remainder of this section, we will find the possible monopole
condensation patterns associated with the three SRO phases.

\subsection{$(0,0,\pi)$ SRO}

\begin{figure}
\includegraphics[width=4cm]{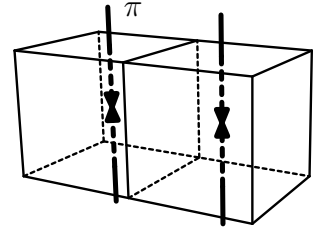}
\caption{\label{fig:flux_00pi} Flux of $\pm\pi$ 
through two of the six plaquettes of the dual lattice. There
is no flux along the directions for which spins are not
staggered.}
\end{figure}

\begin{figure}
\includegraphics[width=5cm]{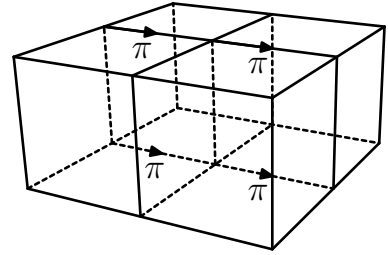}
\caption{\label{fig:gc_00pi} A gauge choice for 
$\widetilde{Y}_{\overline{j}\rho}$
that realizes the fluxes shown in Fig.~\ref{fig:flux_00pi}.}
\end{figure}

To analyze this SRO phase, we first need to find the
monopole hopping amplitudes. To do so we choose an
appropriate gauge (Fig.~\ref{fig:gc_00pi}) that realizes the fluxes on the plaquettes, as shown in 
Fig.~\ref{fig:flux_00pi}. The corresponding monopole hopping 
amplitude is given by
\begin{align}
t_{R,R+\hat{x}}&=t e^{i\pi y}, \notag \\
t_{R,R+\hat{y}}&=t, \notag \\
t_{R,R+\hat{z}}&=l;
\label{hop00pi}
\end{align}
where $t$ and $l$ are real-valued hopping coefficients.
These coefficients emphasize the anisotropic character of the 
monopole hopping due to the presence of diagonal bonds on four faces 
of the cube. After diagonalizing the kinetic part of the 
soft-spin action Eq.~(\ref{sftspnaction}), we find two 
low-energy modes. The normalized real-space wave functions 
associated with these two monopole excitations carrying 
different lattice momenta are
\begin{align}
\Psi_1(R) &= \frac{(1+\sqrt{2})-e^{i\pi y}}{\sqrt{4+2\sqrt{2}}}
\notag \\
\Psi_2(R) &= \frac{((1+\sqrt{2})+e^{i\pi y})e^{i\pi x}}{\sqrt{4+2\sqrt{2}}}
\label{00piwvfct}
\end{align}
To characterize the behavior of this system near monopole
condensation transition, we consider fields which are 
linear combinations of $\Psi_1$ and $\Psi_2$. Any such
linear combination is at the bottom of the monopole band
where there is a continuum of states for the monopoles to
condense:
\begin{equation}
\Phi(R)=\alpha_1 \Psi_1(R) + \alpha_2 \Psi_2(R)
\label{00pilc}
\end{equation}
The phase transition in this system can be explored within
a Ginzburg-Landau theory by treating $\alpha_1$  and $\alpha_2$
as slowly-varying fields. To simplify the next calculations,
we define two new fields in terms of $\alpha_1$ and $\alpha_2$:
\begin{equation}
\phi_1=\alpha_1+i\alpha_2,  \ \ \    \phi_2=\alpha_1-i\alpha_2.
\label{newfields}
\end{equation}
By studying the action of the lattice symmetries, we find that
the resulting Ginzburg-Landau functional is required
to be invariant under the following transformations:
\begin{equation}
\begin{array}{rl} 
T_x: & \phi_1 \rightarrow \phi_2 \\
     & \phi_2 \rightarrow \phi_1 \\
T_y: & \phi_1 \rightarrow i\phi_2 \\ 
     & \phi_2 \rightarrow -i\phi_1 \\ 
T_z: & \phi_1 \rightarrow \phi_1 \\
     & \phi_2 \rightarrow \phi_2 \\
R_{\pi/2,R_{xy}}: & \phi_1 \rightarrow e^{i\pi/4} \phi_1 \\
                  & \phi_2 \rightarrow e^{-i\pi/4} \phi_2 \\
R_{\pi/2,R_{xz}}: & \phi_1 \rightarrow \frac{1}{2} 
\left(\phi_1 + \phi_2 + (\phi_1-\phi_2)(-1)^{x+z}\right) \\ 
                  & \phi_2 \rightarrow \frac{1}{2} 
\left(\phi_1 + \phi_2 - (\phi_1-\phi_2)(-1)^{x+z}\right),
\end{array}
\end{equation}
where the $\pi/2$ rotations are about the lattice points 
on which the monopoles reside. 
These transformations suggest that the simplest invariant takes the form 
\begin{equation}
J(\phi_1,\phi_2)=N_x^2N_y^2,
\end{equation}
with
\begin{equation}
N_{\alpha}(\phi_1,\phi_2)\equiv{\bf \phi^{\dag}
\hat{\sigma}^{\alpha}\phi},
\end{equation}
where $\hat{\sigma}^{\alpha}$ are Pauli matrices.
We then write down the continuum action for the two-component complex
field ${\bf \phi}(R,\tau)$ that respects the above symmetries. In
doing so, we have restored the dual gauge field $L_{\rho}$ and
included some generic kinetic energy $S_L$ for $L_{\rho}$.
\begin{align}
S_{slow}&=\int d\tau d^3R \left(|(\nabla_{\rho}-iL_{\rho}){\bf \phi}|^2
+U(|{\bf \phi}|^2)+wJ({\bf \phi}) \right) \notag \\
&~~~~~+S_L,
\label{sslow}
\end{align}
where $U(|{\bf \phi}|^2)=r|{\bf \phi}|^2+u|{\bf \phi}|^4+\cdots$.
When $r<0$, the monopoles would condense. From now on,
we study confining paramagnetic phases originating from the
Coulomb phase (U(1) spin liquid) by condensing single monopoles.
Ground states are selected by minimizing $wJ({\bf \phi})$, and
the sign of $w$ determines the structure of the resulting phase.

The expectation values $N_x$ and $N_y$ are 
sufficient to characterize each state since the spatial monopole 
density is given by
\begin{equation}
|\Phi(R)|^2=C \left( \frac{1}{2}|{\bf \phi}|^2-(1+\sqrt{2})
((-1)^y N_x+(-1)^x N_y) \right),
\end{equation}
where $C$ is a normalization constant. 
Depending on the sign of $w$, the following
phases are possible.

\subsubsection{$w>0$}
\begin{figure}
\includegraphics[width=4cm]{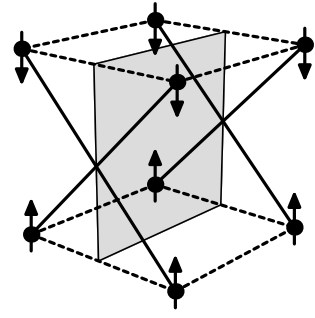}
\caption{\label{fig:col_sl_vbs} Picture of a  ``sheet-like'' columnar 
VBS state obtained when $w>0$. The monopole density is
increased on the x-odd planes which is interpreted as having
antiferromagnetic bonds between NNN spins
preferentially crossing these planes. [Note: dashed lines are
only a guide for the eye.]}
\end{figure}

In this case, there are four ground states
\begin{equation}
(N_x,N_y)=(\pm 1,0),(0,\pm 1).
\label{v8gt0_00pi}
\end{equation}
For each state, the monopole density is the same in every other
plane perpendicular to the $x$ or $y$ lattice axis. For example,
the state $(N_x,N_y)=(0,1)$ has an increased density on the $x$-odd
planes and an decreased density on the $x$-even planes. In the
original spin model, these dual planes are crossed by
antiferromagnetic bonds between
NNN spins, and there is an increased bond energy
crossing the $x$-odd planes. Hence, this state corresponds to a
``sheet-like''   columnar valence bond solid with columns of zigzagging 
dimer chains running along the $z$ direction. These columns are called 
``sheet-like'' because they lie entirely on the $yz$ plane (as illustrated in 
Fig.~\ref{fig:col_sl_vbs}). 

\subsubsection{$w<0$}
\begin{figure}
\includegraphics[width=4cm]{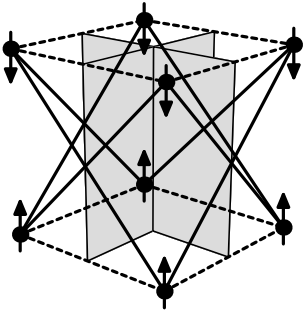}
\caption{\label{fig:col_cl_vbs} Picture of ``cylinder-like'' columnar 
VBS state obtained when $w<0$. The monopole density is increased
on the $x$ and $y$ odd planes which is interpreted
as having antiferromagnetic bonds between NNN spins 
preferentially crossing these planes. [Note: dashed lines are
only a guide for the eye.]}
\end{figure}

In this case, there are four ground states
\begin{equation}
(N_x,N_y)=(\pm 1,\pm 1).
\label{v8st0_00pi}
\end{equation}
For these states, the monopole density oscillates in the $x$ and
$y$ directions. For example, the state $(N_x,N_y)=(1,1)$ has increased
density on odd $x$ and $y$ planes and decreased density on even
$x$ and $y$ planes. In the original spin model, these dual planes are
crossed by 
antiferromagnetic bonds between NNN spins, and there is
an increased bond energy crossing the $x$ and $y$ odd planes.
Consequently, this state corresponds to a ``cylinder-like'' columnar 
valence bond solid. As illustrated in Fig.~\ref{fig:col_cl_vbs}, four 
connecting sheets of zigzagging dimer chains form a cylinder oriented
along the $z$ direction. 

\subsubsection{Connection to the VBS order parameter}
 
Modulation of the monopole density in the dual lattice gives rise to 
the modulation of the dual gauge flux in the dual planes. The modulation of the
dual gauge flux, in turn, corresponds  to the modulation of static electric field 
on the links of the direct lattice. 
Following Ref.~\onlinecite{sachdev89}, we can
identify the valence bond order parameter with these static electric fields.  
The valence bond order parameter in this phase, corresponding to diagonal bonds on the
$xz$ and $yz$ faces of the cubic lattice, and links along the $z$ direction,
can be written as 

\begin{equation}
\vec{\Psi}_{VBS} =
\left(\begin{matrix} (-1)^x (\vec{S}_r \cdot \vec{S}_{r+\hat x+\hat z}+
    \vec{S}_{r+\hat z} \cdot \vec{S}_{r+\hat x})\\
                     (-1)^y (\vec{S}_r \cdot \vec{S}_{r+\hat y+\hat z}+
    \vec{S}_{r+\hat z} \cdot \vec{S}_{r+\hat y}) \\
                     (-1)^z \vec{S}_r \cdot \vec{S}_{r+\hat z}
\end{matrix}\right) ~.
\end{equation}

Using this observation, we can make the following identification,
similar to that in Ref.~\onlinecite{motrunich04}, 
\begin{equation}
\vec{\Psi}_{VBS} \sim \vec{N}=\phi^\dagger\vec{\sigma}\phi .
\end{equation}
Since the diagonal bonds occur in pairs, the perpendicular components of the
spin correlations cancel, and we are left with the spin correlation along
the link directions. Therefore the identification of the diagonal bond order
parameter with the static electric fields in the link may be justified. 

In the above analyses, there is no  modulation of the monopole density in
the $z$-direction; therefore, there is no valence bond solid order
formation in the $z$-direction.  

\subsection{$(0,\pi,\pi)$ SRO}

\begin{figure}
\includegraphics[width=4cm]{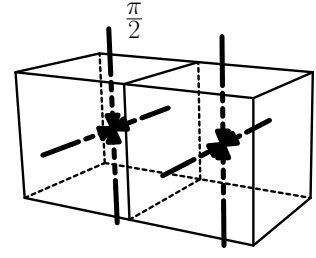}
\caption{\label{fig:flux_0pipi} Flux of $\pm\frac{\pi}{2}$ 
through four of the six plaquettes of the dual lattice. There
is no flux along the direction for which spins are not
staggered.}
\end{figure}

\begin{figure}
\includegraphics[width=5.5cm]{gauge_choice_0pipi.1}
\caption{\label{fig:gc_0pipi} A gauge choice for 
$\widetilde{Y}_{\overline{j}\rho}$
that leads to the fluxes shown in Fig.~\ref{fig:flux_0pipi}.}
\end{figure}

In this phase, upon choosing an appropriate 
gauge (Fig.~\ref{fig:gc_0pipi}), the 
monopole hopping amplitudes (corresponding
to the fluxes on the plaquettes as shown in Fig.~\ref{fig:flux_0pipi})
are given by
\begin{align}
t_{R,R+\hat{x}}&=\frac{t}{\sqrt{2}}\left(e^{i\pi z}+i e^{i\pi y}\right),
\notag \\
t_{R,R+\hat{y}}&=l, \notag \\
t_{R,R+\hat{z}}&=l;
\label{hop0pipi}
\end{align}
where $t$ and $l$ are real-valued hopping coefficients.
\begin{figure}
\includegraphics[height=6cm,width=6cm]{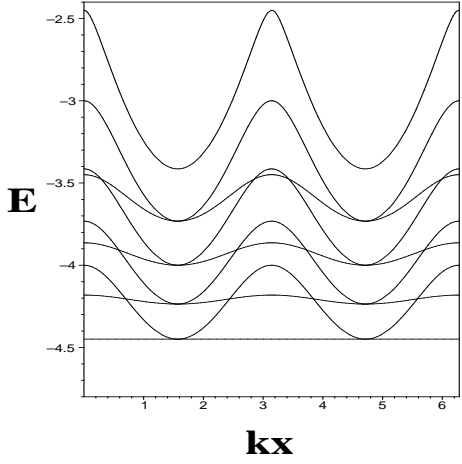}
\caption{\label{kenergy_0pipi} \label{fig:kenergy}
Band structure given by the kinetic energy of the frustrated
XY model in the $(0,\pi,\pi)$ phase. Notice that the
lowest energy band is flat in the $(k_x,0,0)$ direction. 
The second lowest band occurs for the k-vector 
$(k_x,\pi,0)$.}
\end{figure}
After diagonalizing the kinetic part of the soft-spin 
action Eq.~(\ref{sftspnaction}), we find that the lowest energy
band is dispersionless in the $(k_x,0,0)$ direction, as shown in 
Fig.~\ref{fig:kenergy}. This indicates that there is a large degeneracy for the
lowest lying mode at this level of approximation, and no monopole
condensation can occur.  
By going beyond quadratic approximation, this mode may become 
dispersive again and we can then determine the correct monopole 
condensation pattern.
Similar situations have been encountered in the studies of the quantum Ising
model on a Kagome lattice,\cite{nikolic0402,nikolic0403,nikolic03}  and numerical studies
show that a valence bond solid with a larger unit cell is  possibly the 
ground state. A similar structure might also be the ground state, at 
finite $N$, for our $(0,\pi,\pi)$ SRO phase. 
This situation, however, could also be an indication that, at
finite $N$, the spin liquid is more robust in the case of $(0,\pi,\pi)$ SRO
phase than it is for the other two SRO phases found in our study.  Further studies are 
needed to properly determine the true ground state for this phase.  

\subsection{$(\pi,\pi,\pi)$ SRO}

\begin{figure}
\includegraphics[width=4.1cm]{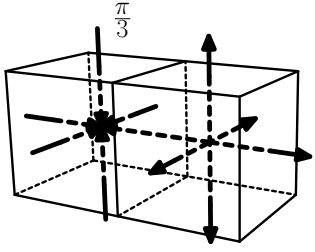}
\caption{\label{fig:flux_pipipi} Flux of $\pm\frac{\pi}{3}$ 
through the six plaquettes of the dual lattice.}
\end{figure}

\begin{figure}
\includegraphics[width=5.5cm]{gauge_choice_pipipi.1}
\caption{\label{fig:gc_pipipi} A gauge choice for 
$\widetilde{Y}_{\overline{j}\rho}$
that realizes the fluxes shown in Fig.~\ref{fig:flux_pipipi}.}
\end{figure}

Repeating the same procedure, we first find the monopole 
hopping amplitudes (corresponding to the fluxes on the 
plaquettes as shown in Fig.~\ref{fig:flux_pipipi}). 
Upon choosing the appropriate
gauge (Fig.~\ref{fig:gc_pipipi}), these amplitudes are given by
\begin{eqnarray}
t_{R, R+\hat{x}} &=& \sqrt{\frac{3}{8}} \Big(1+ie^{i\pi(x+y)}\Big)
+ \sqrt{\frac{1}{8}} \Big(1-ie^{i\pi(x+y)}\Big) e^{i\pi z}, \notag \\
t_{R, R+\hat{y}} &=& \sqrt{\frac{3}{8}} \Big(1-ie^{i\pi(x+y)}\Big)
+ \sqrt{\frac{1}{8}} \Big(1+ie^{i\pi(x+y)}\Big) e^{i\pi z}, \notag \\
t_{R, R+\hat{z}} &=& 1.
\end{eqnarray}
Similar monopole hopping problem was studied in the 
context of an interacting boson model.\cite{motrunich04} As such,
our analysis which follows shares some similarities  with the discussions in 
the bosonic model,\cite{motrunich04} and for completeness, we present
relevant details of the analysis. 
 
We diagonalize the kinetic energy and find two low-energy modes
\begin{eqnarray}
\Psi_1(R) &=& \frac{1 + (\sqrt{3}-\sqrt{2}) e^{i\pi z}}
                         {\sqrt{2(3-\sqrt{6})}}, \notag \\
\Psi_2(R) &=& \frac{1 - (\sqrt{3}-\sqrt{2}) e^{i\pi z}}
                         {\sqrt{2(3-\sqrt{6})}} \times
\frac{e^{i\pi x} - i e^{i\pi y}}{\sqrt{2}}.
\end{eqnarray}
Any linear combination
of these two monopole excitations, as Eq.~(\ref{lincomb}),
is at the bottom of the monopole band. Hence, there is a
continuum of states for the monopoles to condense.
\begin{equation}
\Phi(R) = \phi_1 \Psi_1(R) + \phi_2 \Psi_2(R),
\label{lincomb}
\end{equation}
Near monopole condensation, we can analyze the transition using the 
Ginzburg-Landau formalism. By studying the action of the lattice symmetries, 
we notice that the resulting Ginzburg-Landau 
functional is required to be invariant under the following transformations:
\begin{equation}
\begin{array}{rll}
T_x: & \phi_1 \to \phi_1^* ~~& \phi_2 \to -\phi_2^*\\
T_y: & \phi_1 \to \phi_1^* ~~& \phi_2 \to \phi_2^* \\
T_z: & \phi_1 \to \phi_2^* ~~& \phi_2 \to \phi_1^* \\
R_{\pi/2, Rxy}: & \phi_1 \to e^{-i\pi/4} \phi_1^*
                ~~&\phi_2 \to e^{i\pi/4} \phi_2^* \\
R_{\pi/2, Rxz}: & \phi_1 \to \frac{\phi_1^* + \phi_2^*}{\sqrt{2}}
                ~~&\phi_2 \to \frac{\phi_1^* - \phi_2^*}{\sqrt{2}}.
\end{array}
\end{equation}
It appears from these transformations that the simplest 
invariant has the form 
\begin{equation}
K(\phi_1,\phi_2)=N_x^2N_y^2+N_y^2N_z^2+N_z^2N_x^2,
\end{equation}
with
\begin{equation}
N_{\alpha}(\phi_1,\phi_2)\equiv{\bf \phi^{\dag}
\hat{\sigma}^{\alpha}\phi}.
\end{equation}
We then write down the continuum action for the two-component complex
field ${\bf \phi}(R,\tau)$ that respects the above symmetries. While
doing so, we restore the dual gauge field $L_{\rho}$ and
include some generic kinetic energy for this field. The resulting action 
has the same form as the continuum action presented for the $(0,0,\pi)$
SRO phase (Eq.~(\ref{sslow})), and, as before, the ground states are selected 
by minimizing $w K({\bf \phi})$.

The expectation values $N_x$, $N_y$ and $N_z$ are sufficient to 
characterize each state since the spatial monopole density 
is given by
\begin{equation}
|\Phi(R)|^2= |{\bf \phi}|^2 + \frac{1}{\sqrt{3}}\Big[
(-1)^x N_x+(-1)^y N_y+(-1)^z N_z \Big].
\end{equation}
Then, according to the sign of $w$, we can find which 
valence bond solid structures are allowed.

\subsubsection{$w>0$}
\begin{figure}
\includegraphics[width=4cm]{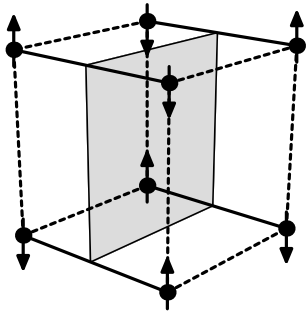}
\caption{\label{fig:col_val_pipipi} Picture of a columnar
VBS state obtained when $w>0$. The monopole density is increased
on the x-even planes which is interpreted as having antiferromagnetic
bonds between NN spins preferentially crossing these
planes. [Note: dashed lines are only a guide for the eye.]}
\end{figure}

In this case, there are six ground states
\begin{equation}
(N_x,N_y,N_z)=(\pm 1,0,0),~~(0,\pm 1, 0),~~(0,0,\pm 1).
\label{v8gt0_pipipi}
\end{equation}
In a given state, the monopole density is the same in every
other plane perpendicular to a fixed lattice axis. For 
example the state $(N_x,N_y,N_z) = (1,0,0)$ has an increased
density on the $x$-even planes and decreased density on the
$x$-odd planes, as shown in Fig.~~\ref{fig:col_val_pipipi}. 
In the original spin model, these dual planes are crossed by antiferromagnetic
bonds between NN spins, and there is an increased
bond energy crossing the $x$-even planes. Hence, this state
corresponds to a columnar valence bond solid with dimers 
oriented in the $x$ direction.

\begin{figure}
\includegraphics[width=4cm]{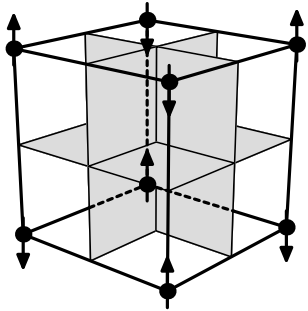}
\caption{\label{fig:box_val_pipipi}Picture of a 3D box VBS state
obtained when $w<0$. The monopole density is increased on all
$x$, $y$ and $z$ even planes, and the dimers resonate around cubes
center where these three planes meet. [Note: dashed lines are parts of the
real bonds.]}
\end{figure}

\subsubsection{$w<0$}
For this second case, there are eight ground states
\begin{equation}
(N_x,N_y,N_z)= \frac{1}{\sqrt{3}}(\pm 1,\pm 1,\pm 1)
\label{v8st0_pipipi}
\end{equation}
where each of the three signs can be chosen independently.
The monopole density for these states is periodic in all three 
directions and correspond to three-dimensional ``box'' valence 
bond solids. For example, the state $(N_x, N_y, N_z) = 
\frac{1}{\sqrt{3}} (1, 1, 1)$ has maximal monopole density 
for even $x$, $y$ and $z$, as shown in Fig.~\ref{fig:box_val_pipipi}.
In the original spin model, this state has dimers resonating
around direct lattice cubes surrounding the dual lattice points.


\section{Conclusion \label{conclusion}}

In this paper, we studied the quantum ground states of the Sp($N$)
antiferromagnetic Heisenberg model on the cubic lattice with NN ($J_1$),
and NNN ($J_2$) exchange interactions. This is the simplest 3D model with frustration. 
We have shown that the sole addition of the NNN interaction on the cubic lattice gave
rise to three neighboring U(1) spin liquids with 
magnetic short-range correlations at $(\pi,\pi,\pi), (0,\pi,\pi), (0,0,\pi)$. 
Moreover, there are indications that $(0,\pi,\pi)$ U(1) spin liquid is more 
likely to ``survive'' at finite $N$ values. 
This model has an advantage over other models because of its direct
connection to realistic microscopic models. Thus, verifying the results
of our model in real systems would be more amenable.

We first obtained the mean-field
phase diagram of this model in the $N\rightarrow\infty$ limit as a function
of $J_2/(J_1+J_2)$, which controls the frustration, and $1/\kappa$, which 
controls quantum fluctuations. For large values of $\kappa$, we find various 
magnetically ordered phases;  $(\pi,\pi,\pi)$, $(0,\pi,\pi)$, $(0,0,\pi)$ collinear-ordered 
and $(q,\pi,\pi)$, $(0,q,\pi)$ helically-ordered states. 
Upon decreasing $\kappa$, we find
three quantum-disordered paramagnetic phases (SRO) with
enhanced spin correlations at $(\pi,\pi,\pi), (0,\pi,\pi), (0,0,\pi)$.
Since these are quantum-disordered states of collinear-ordered magnets,
we see already at the mean-field level that these phases are possible 
U(1) spin liquids.

Methods have been developed by Haldane,\cite{haldane88} Read, and 
Sachdev\cite{sachdev89} to study (2+1)D quantum antiferromagnets. In this work,
we have extended these methods to (3+1)D  and analyzed the effects 
of singular fluctuations about the mean-field
states. These monopole events, or the topological defects in the U(1) gauge theory,
represent tunneling between different topologically distinct sectors.  
We find that, when $N\rightarrow\infty$ (small monopole
fugacity), the three paramagnetic phases are indeed U(1) spin liquids.
At finite $N$ values (large monopole fugacity ), 
the system may become confined due to monopole condensation,
and we discover various possible VBS ordered phases for 
SRO $(\pi,\pi,\pi)$ and $(0,0,\pi)$ states. 
However, it appears that, when $J_1\approx J_2$, the corresponding
U(1) spin liquid with $(0,\pi,\pi)$ SRO seems to be more stable in the 
simplest consideration of monopole condensation than the two other
spin liquid states. We would need a more elaborate analysis to determine
the ultimate fate of this state. One possibility would be 
that it becomes a VBS state with a very large unit cell.\cite{nikolic03,nikolic0402}
Another possibility is that it remains a U(1) spin liquid 
near $J_1 \approx J_2$ even at finite $N$.
If the latter is the case, this U(1) spin liquid may be a better
candidate to be observed in real systems.
One may be able to gain more insights on this issue using 
numerical methods.

It is worthwhile to mention that the U(1) spin liquids studied previously,
in the context of other three-dimensional models, correspond to 
our U(1) spin liquid with SRO $(\pi,\pi,\pi)$.\cite{huse04,hermele04,motrunich04,tchern04} 
On the other hand, 
we find two other possible U(1) spin liquids with $(0,\pi,\pi)$ and 
$(0,0,\pi)$ correlations within our model.
It is also interesting to note that our analysis for the quantum-disordered 
paramagnetic phases at finite $N$ shares some similarities with 
the studies of an interacting boson system at half-filling.\cite{motrunich04}
The nature of the phase transitions between different U(1) spin liquid
states and VBS phases is still not clear and awaits further investigation.

\begin{acknowledgments}
This work was supported by NSERC of Canada, Canada Research Chair
program, Canadian Institute for Advanced Research  (YBK) and FQRNT-Qu\'ebec (JSB). 
We would also like to thank O. Motrunich, T. Senthil, 
and M.P.A. Fisher for helpful discussions, and S. Takei for his 
technical assistance. YBK is grateful to Kavli Institute for Theoretical Physics
and Aspen Center for Physics for their hospitality where some parts of 
this work were performed. 
\end{acknowledgments}

\end{document}